# High-order minibands and interband Landau level reconstruction in graphene moiré superlattice

Xiaobo Lu[1,2†*], Jian Tang[1†], John R. Wallbank[3], Shuopei Wang[1], Cheng Shen[1], Shuang Wu[1], Peng Chen[1], Wei Yang[1,2], Jing Zhang[1], Kenji Watanabe[4], Takashi Taniguchi[4], Rong Yang[1], Dongxia Shi[1], Dmitri K. Efetov[2], Vladimir I. Fal′ko[3], Guangyu Zhang[1,5,6*]

[1] Beijing National Laboratory for Condensed Matter Physics and Institute of Physics, Chinese Academy of Sciences; CAS Key Laboratory of Nanoscale Physics and Devices, Beijing 100190, China

[2] ICFO - Institut de Ciencies Fotoniques, The Barcelona Institute of Science and Technology, 08860 Castelldefels, Barcelona, Spain

[3] National Graphene Institute, University of Manchester, Manchester M13 9PL, UK

[4] National Institute for Materials Science, 1-1 Namiki, Tsukuba, 305-0044, Japan

[5] Collaborative Innovation Center of Quantum Matter, Beijing 100190, China

[6] Songshan-Lake Materials Laboratory, Dongguan 523808, Guangdong Province, China

†Equal contribution

**The propagation of Dirac fermions in graphene through a long-period periodic potential would result in a band folding together with the emergence of a series of cloned Dirac points (DPs)[1,2]. In highly aligned graphene/hexagonal boron nitride (G/hBN) heterostructures, the lattice mismatch between the two atomic crystals generates a unique kind of periodic structure known as a moiré superlattice. Of particular interests is the emergent phenomena related to the reconstructed band-structure of graphene, such as the Hofstadter butterfly[3-5], topological currents[6], gate-dependent pseudospin mixing[7], and ballistic miniband conduction[8]. However, most studies so far have been limited to the lower-order minibands, e.g. the 1$^{st}$ and 2$^{nd}$ minibands counted from charge neutrality, and consequently the fundamental nature of the reconstructed higher-order miniband spectra still remains largely unknown. Here we report on probing the higher-order minibands of precisely aligned graphene moiré superlattices by transport spectroscopy. Using dual electrostatic gating, the edges of these high-order minibands, i.e. the 3$^{rd}$ and 4$^{th}$ minibands, can be reached. Interestingly, we have observed interband Landau level (LL) crossing**

**inducing gap closures in a multiband magneto-transport regime, which originates from band overlap between the 2$^{nd}$ and 3$^{rd}$ minibands. As observed high-order minibands and LL reconstruction qualitatively match our simulated results. Our findings highlight the synergistic effect of minibands in transport, thus presenting a new opportunity for graphene electronic devices.**

The quantum nature of electrons in a spatially periodic lattice generates a band structure. Similarly, electrons in crystalline solids moving through longer-period superlattice develop a finer energy spectrum consisting of discrete minibands. Manipulating electronic states by superlattices to overcome the constraints of the original materials has been a remarkable route for band engineering. The moiré superlattice of G/hBN has emerged recently as a model two dimensional (2D) system, which can effectively fold the band spectrum without destroying the original honeycomb lattice of graphene. In contrast to conventional mesoscale scalar potentials, the moiré potential arises from atomic scale potentials in hBN and should be described by a two-by-two tensor[9,10], leading to a much more complicated band spectrum. According to previous simulations[10], different minibands can overlap with each other and obscure the presence of extra Dirac points. So far, most attention has been focused on the second-generation Dirac cones (between the 1st and 2nd minibands). However, the exploration of higher-order minibands has been much more limited. Hence, the study of high-order minibands in G/hBN moiré superlattice is crucial to fully understand the emergent quantum behavior of Dirac fermions in a periodic spinor potential.

Probing high-order minibands of G/hBN moiré superlattice has been experimentally challenging. The modulation of the chemical potential by electrostatic gating is usually not sufficient to reach the edges of high-order minibands due to the breakdown of conventional solid gate dielectrics (hBN/SiO$_2$). Besides, the band spectrum of the moiré superlattice depends very sensitively on the twist angle between graphene and hBN[10,11]. The density of electron states of each completely filled miniband can be described as $n_0 = 1/A$ (if not counting spin-valley degeneracy), where $A = \sqrt{3}\lambda^2/2$ is the area of a single moiré unit cell and $\lambda$ is the moiré wavelength. When two crystals are precisely aligned, it gives the maximum moiré wavelength $\lambda = \sim 15$ nm[12-14]. When the twist angle increases, $\lambda$ decreases drastically, resulting in a repaid increase of the carrier capacity for each miniband. As a result, the perfectly aligned G/hBN moiré superlattice has lowest carrier density for each filled miniband and is an ideal system for probing high-order minibands.

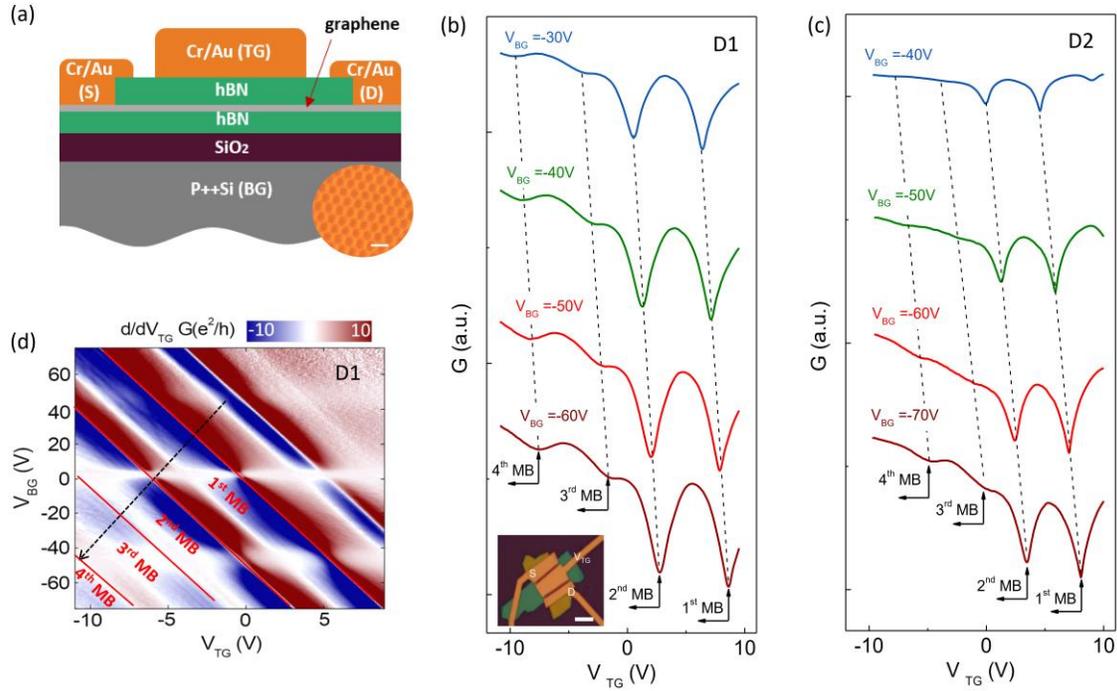

**Fig.1. Dual-gated G/hBN moiré superlattice device.** (**a**) Schematic cross-sectional view of device. Insert is a high-resolution AFM image of precisely aligned G/hBN moiré superlattices, showing a well-defined moiré pattern with a period of ~15 nm. The scale bar is 20 nm. (**b**) (**c**) Conductance of two independent devices as a function of top gate voltage $V_{TG}$ at different bottom gate voltage $V_{BG}$ with different minibands edges indicated by dashed lines. Curves at different $V_{BG}$ have been off-set for clarity. Insert in (**b**) shows an optical microscopy image of an as fabricated device with scale bar 10 um. (**d**) Color plot of $dG/dV_{TG}$ as a function of $V_{TG}$ and $V_{BG}$ measured from devices D1. All measurements are done at a base temperature of 1.5K.

In this study, we present magneto-transport studies of zero-twisted G/hBN moiré superlattices. Fig. 1a shows a schematic of the device structure in which dual-gate geometry is adopted. Samples were prepared by a van der Waals epitaxy technique[13], followed by stacking of another hBN flake on top of it (by dry-transfer technique) as top gate dielectric. A high resolution atomic force microscopy (AFM) image of the epitaxial graphene on hBN is displayed in Fig. 1a (insert), showing a well-defined moiré pattern with a period of λ=~15 nm. Cr/Pd/Au (1nm/10nm/100nm) was then deposited for contacts and gate electrodes. Fig. 1b-c show the typical top gate dependent conductance measured from two independent devices at different bottom gate values. Original DP and second-generation DP (SDP) are reflect on the minimums of conductance which are marked by onsets of the first miniband (1$^{st}$ MB) and the second miniband (2$^{nd}$ MB), respectively. Interestingly, at higher carrier density we find two more conductivity minimums, most likely originated from high-order minibands. Fig. 2d shows the numerical derivative of the conductance $dG/dV_{TG}$ as a function of $V_{TG}$ and $V_{BG}$. The minimums of conductivity are further resolved in Fig. 2d (red lines). Some other features at $V_{BG} = 45\ V, 5V$ and $-35V$ are also observed and expected to come from non-top-gated area of graphene, since they are independent on $V_{TG}$. Considering the spin and valley degeneracy in graphene, the capacity of each miniband is $4n_0$. According to previous studies, second-generation Dirac points (SDPs) are always locates at a carrier density of $n = 4n_0$ where $n_0$ only depends on the twist angle. For precisely aligned G/hBN moiré superlattices[13,14], that $4n_0 = 1.85 \times 10^{12} cm^{-2}$. Thus, by employing the parallel plate capacitor model, we can deduce the carrier density at any gate voltages. To keep simplicity, we use to $n_0$ normalize the carrier density in below ($0, -4n_0, -6.5n_0$ and $-11.2n_0$ are the normalized onset carrier density of 1$^{st}$ MB, 2$^{nd}$ MB, 3$^{rd}$ MB and 4$^{th}$ MB, respectively). Note that, both two devices with top layer hBN randomly stacked show same features at same carrier density which can be normalized by that of SDP, thus ruling out the possibility of super moiré induced by top layer hBN.

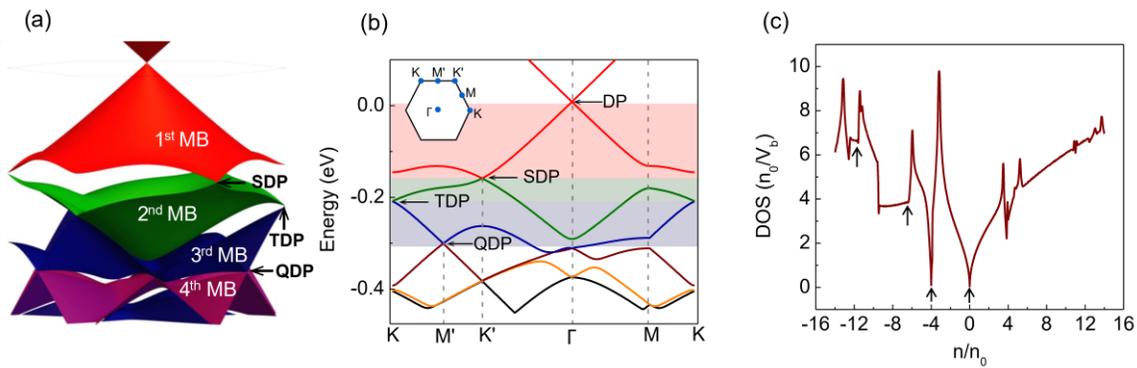

Fig.2. **High-order minibands (MB) in G/hBN moiré superlattices.** (**a**) Calculated band spectrum in the SBZ, with different minibands indicated by different color. (**b**) Calculated energy dispersions along the $K - M' - K' - \Gamma - M - K$ directions of SBZ shown in the insert figure.

(c) Calculated density of states (DOS). Positions corresponding to four different DPs are indicated by black arrows.

To gain a better understanding of these results, we thus calculated the band structure of a precisely aligned G/hBN moiré superlattice using the general symmetry-based approach described in a previous work[8,10]. The superlattice parameters were extracted from Ref. 8. Fig. 2a displays the band spectrum in the superlattice Brillouin zone (SBZ) with different minibands rendered with different colors. The points of contact between different minibands occur at SDPs[12], tertiary Dirac points (TDPs)[15], and quaternary Dirac points (QDPs), all found at the edges of the SBZ. Fig. 2b shows the energy dispersion along the K-M'-K'-Γ-M-K direction in the SBZ from which the four generations of DPs can also be identified. The SDP is located at the K' point of the SBZ at energy −170 meV relative to the original DP, which is consistent with the ARPES results[16]. TDPs and QDPs are also marked in Fig. 2a and 2b. According to the simulated dispersions, TDPs are located at the K point in the SBZ at energy −210 meV below the original DP while the QDPs are located at the M' points and −300 meV. Fig. 2c shows the density of state (DOS) plotted against normalized carrier densities. When the carrier density is gradually tuned to $-6.5n_0$, the Fermi surface starts to touch the TDP which is also the onset point of the third miniband. An observed increase of sample resistivity can thus be attributed to the abrupt decrease of the DOS as well as the opening of an additional interband scattering channel[17-19]. Similarly, when the carrier density is further tuned to $-11.2n_0$, the Fermi surface starts to touch QDPs and results in another abrupt increase of resistivity.

Under a magnetic field, the energy spectrum of a graphene superlattice is reconstructed into sequences of magnetic Bloch bands and LLs[20,21]. The plot of the correspondingly reconstructed carrier density against the magnetic field, known as a Wannier diagram, has its incompressible states described by the following Diophantine relation[21]

$$(n/n_0) = t(\Phi/\Phi_0) + s$$

where n is the carrier density; $n_0$ is the density of electron states of each Bloch band; $\Phi$ is the magnetic flux per moiré unit cell; $\Phi_0 = h/e$ is the magnetic flux quantum; t is LL filling index and s is Bloch band filling index. Here both t and s should be integer.

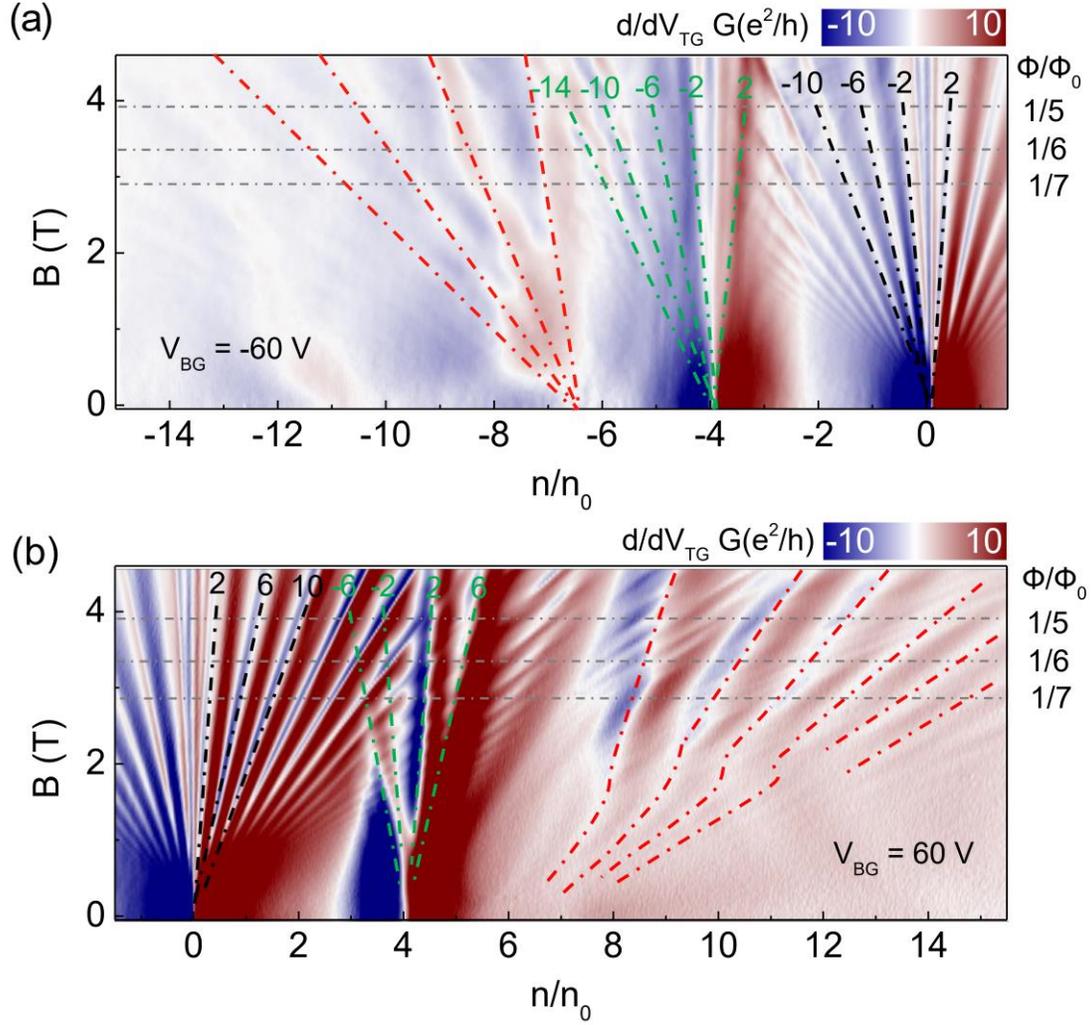

**Fig. 3. Anomalous Quantum Hall states in G/hBN moiré superlattice.** (**a**) (**b**) dG/dn as a function of normalized carrier density n/n$_0$ and magnetic field B with V$_{BG}$ set at (**c**) −60 V and (**d**) 60V, respectively. Data is measured from device D1.

Fig. 3a-b show the differential conductivity dG/dn as a function of normalized carrier densities n/n$_0$ and the magnetic field B with bottom gate voltage V$_{BG}$ set at −60 V and 60V respectively. Landau fans tracing from the DPs and SDPs strictly follow the rule described in the Diophantine relation. For LLs fanning out from Dirac point, s = 0 and t = 4(N + 1/2) where N = 0, ±1, ±2 etc, shows features of typical half integer quantum hall effect in monolayer graphene. For LLs fanning out from SDPs (n = ±4n$_0$), fitting of Diophantine relation gives s = 4 and t = 4(N + 1/2) where N = 0, ±1, ±2 etc. Here, both Bloch bands and Landau levels are four fold degenerated due to spin and valley degrees of freedom. Interestingly, at higher carrier densities,

we observed extra sets of fan diagrams, which are marked by red dashed lines in Fig. 3a-b. As we can see that, the fan diagrams are no longer symmetric and only features tracing to one side are observable. We note that in non-perfectly aligned G/hBN moiré superlattice, the LL-like signitures fanning out from TDP hosts a degeneracy of twelve, which was attributed to a $\sqrt{3}\times\sqrt{3}$ R 30° hidden superstructure[15]. Here we also find the spacing between adjacent features faning out from TDP is much larger than those found for the DP and SDP, however, still smaller than twelve (close to eight). On the electron side, the LL-like features are not even straight. As-observed the features cannot be well described by Diophantine relation and are most likely induced by interband LLs crossing.

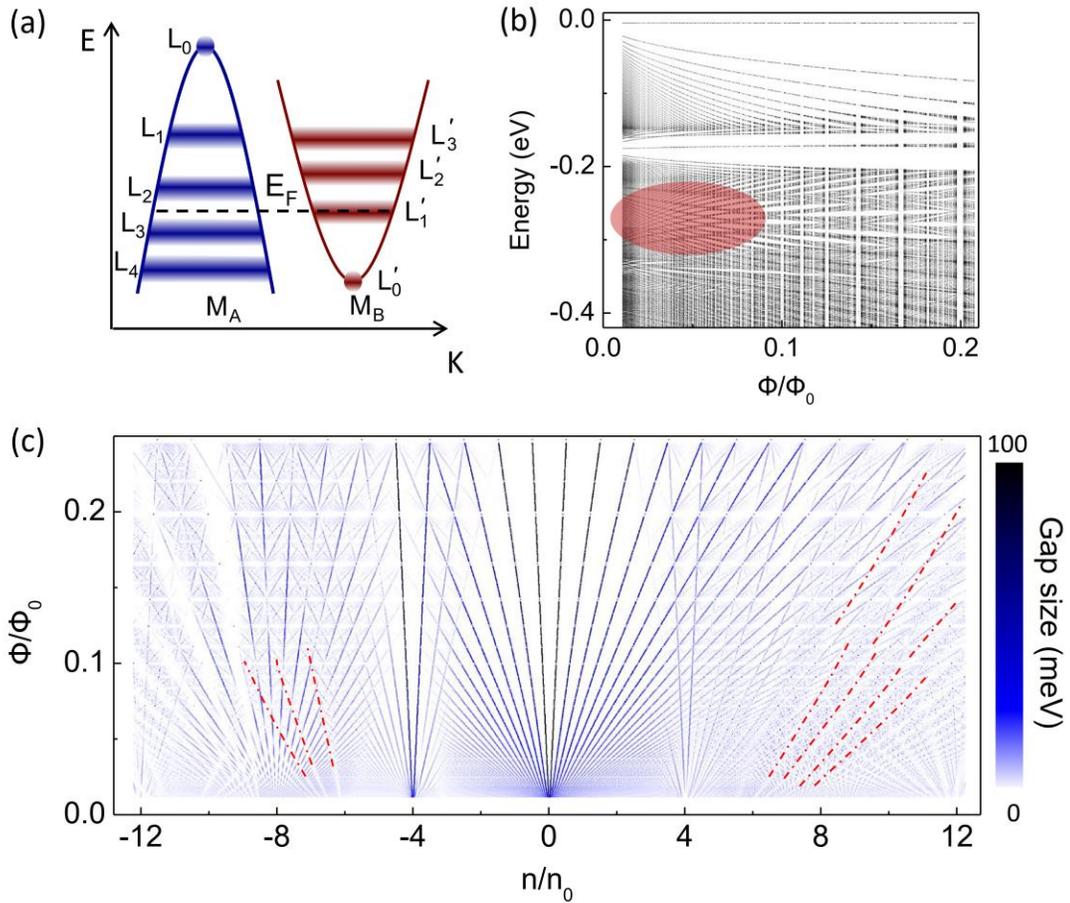

Fig. 4. **Interband LL reconstruction induced gap closing.** (**a**) Schematic showing interband LL crossing induced gap closing. (**b**) Numerically calculated LL spectrum on the hole side. A region of crossing LLs is indicated with red shading. (**c**) Calculated Wannier diagram of a precisely aligned graphene moiré superlattice with color rendering corresponding to the different gap sizes. The red dashed lines indicate the areas of LL crossing induced gap closing.

In a multiband transport regime, LLs from overlapping minibands with different effective electron masses cross with each other at some finite magnetic field, leading to a reconstructed energy spectrum[22]. As shown in Fig. 4a, the gap between two neighboring LLs, e.g. $L_2$ and $L_3$, from miniband A ($M_A$) is filled by $L_1$ from miniband B ($M_B$). As a result, the sample will be conductive even without edge state when the Fermi surface is tuned inside the gap between $L_2$ and $L_3$ in $M_A$. According to Fig. 2b, there will be a transition from single-band transport to multiband transport in precisely aligned G/hBN moiré superlattice when the Fermi surface is tuned to TDP at ~-210 meV and thus makes it possible to observe the induced gap closing due to interband LLs crossing in a multiband transport regime. Fig. 4b displays the calculated the energy spectrum in magnetic field which is focused on the hole side. LLs traced to two overlapping minibands produce a criss-crossed pattern which is visible in the area marked by the red shading.

To investigate these LL crossing features, we further calculated the density-magnetic field diagram in which different colors are used to display the different gap sizes (Fig.4c). In a single particle picture, LL gaps in n-B plain always trace back to an integer miniband filling factor at $B = 0$ (e.g. $n = 0, \pm 4n_0, \pm 8n_0 ...$)[23], which is also clearly shown in Fig. 4c. Qualitatively consistent with experimental the results, interband LL crossing in a multiband transport regime, leads to sequences of gap closures (indicated by the red dashed lines in Fig. 4c) which are traced to non-integer band filling at $B = 0$. The winding features on the electron side (indicated by red dashed lines in Fig. 2b) are also clearly resolved.

According to our simulated results, the interband LL crossing induced gap closing in hole side is only visible at low magnetic fields $B < 2T$. The value is lower compared experimental value (~4T). Meanwhile, the calculation cannot well interpret the spacing between adjacent features (almost two times of the spacing for four-fold LLs) fanning out from TDP on the hole side. These discrepancies may come from the non-negligible role of interactions in the moiré system, especially at high carrier densities.

In conclusion, we carried out transport measurements to probe the minibands of precisely aligned graphene moiré superlattices at high carrier densities. We observed TDPs and QDPs at carrier densities of ∼6.5$n_0$ and ∼11.2 $n_0$, respectively, which manifest as an abrupt increase of sample resistivity. Gap closing induced by interband LL crossing is observed and can be explained by a multiband transport mechanism. Our results reveal a clear picture of the band structure of perfectly aligned G/hBN moiré superlattices and indicate the important role of the synergistic effect of minibands in transport.

■AUTHOR INFORMATION

Corresponding Author
*E-mail: Xiaobo.Lu@icfo.eu; gyzhang@iphy.ac.cn

Author Contributions

G.Z. and X.L. designed the research; X.L. and J.T. performed the growth, structural characterization, device fabrication, and electrical transport measurements; J.W. and V.F. performed the calculation. K.W. and T.T. synthesized h-BN crystals. X.L., J.W., G.Z. and V. F. analyzed data and wrote the manuscript, and all authors discussed and commented on the paper.

Notes

The authors declare no competing financial interest.

■ACKNOWLEDGMENTS
G.Z. thanks the National Science Foundation of China (NSFC) under Grant No. 11834017, National Key R&D program under Grant No. 2016YFA0300904, the Strategic Priority Research Program of Chinese Academy of Sciences (CAS) under Grant No. XDB30000000, and the Key Research Program of Frontier Sciences of the CAS under Grant No. QYZDB-SSW-SLH004. K.W. and T.T. acknowledge support from the Elemental Strategy Initiative conducted by the MEXT, Japan and and the CREST (JPMJCR15F3), JST.